\documentclass[onecolumn,%
               amsmath,%
               amssymb,%
               amsfonts,%
               superscriptaddress,%
               nofootinbib,%
               notitlepage,%
               showpacs,%
               aps,%
               prd,%
               10pt,%
               ]{revtex4-1}
\usepackage[utf8x]{inputenc}
\usepackage[english]{babel}
\usepackage{fontenc,%
            graphicx,%
            color,%
            }
\usepackage[hidelinks]{hyperref}

\def\be{\begin{equation}}
\def\ee{\end{equation}}

\allowbreak
\allowdisplaybreaks

\begin{document}

\title{Dark Energy and Dark Matter from an additional adiabatic fluid}

\author{Peter K. S. Dunsby}
\email{peter.dunsby@uct.ac.za}
\affiliation{Department of Mathematics and Applied Mathematics, University of Cape Town, Rondebosch 7701, Cape Town, South Africa.}
\affiliation{Astrophysics, Cosmology and Gravity Centre (ACGC), University of Cape Town, Rondebosch 7701, Cape Town, South Africa.}
\affiliation{South African Astronomical Observatory, Observatory 7925, Cape Town, South Africa.}

\author{Orlando Luongo}
\email{luongo@na.infn.it,\,\,luongo@uct.ac.za}
\affiliation{Department of Mathematics and Applied Mathematics, University of Cape Town, Rondebosch 7701, Cape Town, South Africa.}
\affiliation{Astrophysics, Cosmology and Gravity Centre (ACGC), University of Cape Town, Rondebosch 7701, Cape Town, South Africa.}
\affiliation{Dipartimento di Fisica, Universit\`a di Napoli ''Federico II'', Via Cinthia, Napoli, Italy.}
\affiliation{Istituto Nazionale di Fisica Nucleare (INFN), Sez. di Napoli, Via Cinthia, Napoli, Italy.}

\author{Lorenzo Reverberi}
\email{lorenzo.reverberi@uct.ac.za}
\affiliation{Department of Mathematics and Applied Mathematics, University of Cape Town, Rondebosch 7701, Cape Town, South Africa.}
\affiliation{Astrophysics, Cosmology and Gravity Centre (ACGC), University of Cape Town, Rondebosch 7701, Cape Town, South Africa.}

\begin{abstract}
The Dark Sector is described by an \emph{additional} barotropic fluid which evolves adiabatically during the universe's history and whose
adiabatic exponent $\gamma$ is derived from the standard definitions of specific heats. Although in general $\gamma$ is a
function of the redshift, the Hubble parameter and its derivatives, we find that our assumptions lead necessarily to solutions
with $\gamma = $ constant in a FLRW universe. The adiabatic fluid acts effectively as the sum of two distinct components, one
evolving like non-relativistic matter and the other depending on the value of the adiabatic index.  This makes the model
particularly interesting as a way of simultaneously explaining the nature of both Dark Energy and Dark Matter, at least at the
level of the background cosmology. The $\Lambda$CDM model is included in this family of theories when $\gamma = 0$. We fit our
model to SNIa, $H(z)$ and BAO data, discussing the model selection criteria. The implications for the early-universe and the growth
of small perturbations in this model are also discussed.
\end{abstract}


\date{\today}

\maketitle

\section{Introduction}

The present time cosmic expansion may be described in terms of a late-time fluid which dominates over the other contributions to the cosmic matter budget
\cite{galaxy}. The simplest assumption is based on the hypothesis that such a fluid is \emph{perfect} \cite{bamba}
and enters \emph{by hand} the Einstein equations as source for speeding up the universe today \cite{adam}. This component
(referred to as ``Dark Energy") is required to have a negative equation of state in order to guarantee that the universe undergoes an
accelerated phase at late times \cite{negativep}, and the search for its nature is the focus of much current research in cosmology.
The \emph{minimal model} for Dark Energy is the one where the cosmological constant $\Lambda$ \cite{lambda1} dominates over the other
species including pressureless matter \cite{lambda2}. Although appealing and now very well-established, the cosmological constant
suffers from several shortcomings and consequently the $\Lambda$CDM model cannot be considered the complete explanation for the universe dynamics\footnote{All conceivable approaches to Dark Energy are practically indistinguishable at the level of
the background, leading to a degeneracy problem and only model-independent measures of the evolution of the equation of state would
indicate whether the $\Lambda$CDM model really is the favoured cosmological framework \cite{emay12}.}\cite{lambda3}.

It is for these reasons that models for an evolving Dark Energy contribution have attracted considerable attention over the past two
decades \cite{eoevolve}. There exist several explanations for \emph{evolving Dark Energy}, which range from modifying Einstein's gravity,
including additional degrees of freedom arising from quantum backgrounds to proposing different energy momentum tensors for this dark
sector\footnote{For a representative but incomplete list, see for example \cite{revcop} and references therein.}. In every cases, all
evolving Dark Energy contributions should be compatible with the laws of thermodynamics and be described by a perfect fluids, at
least at the level of the background cosmology \cite{thermodynamic}. The problem of describing properties of equilibrium
thermodynamics in terms of a non-equilibrium Dark Energy fluid is a one of the challenges of modern day cosmology \cite{revtermo}.

Within the framework of a homogeneous and isotropic universe, this problem can be avoided by assuming that at any given epoch, the
fluid evolution is at least described by a quasi-static process. More recently, it has been shown that it is possible
to formulate the thermodynamic quantities of interest for a Friedmann-Lema\^itre-Robertson-Walker (FLRW) cosmology
\cite{razi1}. In particular, it has been argued that the role of specific heats in cosmology can be confronted with
observational data \cite{altro}. In doing so, an investigation of the simplest assumptions on specific
heats leading to an evolving Dark Energy contribution has been recently presented \cite{ioequevedo}.

In this paper we show that by investigating how heat capacities evolve at arbitrary redshift it may be possible to construct a
cosmological model with an evolving Dark Energy term, which is a natural extension to the standard $\Lambda$CDM model.

In what follows, a few basic requirements for the heat capacities are assumed:
\begin{enumerate}
\item they evolve in time,
\item they are related to the internal energy and enthalpy of the universe as required by standard thermodynamics,
\item they have been evaluated for all perfect fluids making up the energy budget of the universe,
\item the process of thermal exchange is purely adiabatic, so that the volume scales as the third power of the scale factor.
\end{enumerate}
Moreover, as a consequence of the above prescriptions, the pressureless matter contribution turns out to be an emergent
phenomenon and that the inferred Dark Energy contribution is weakly interacting, behaving like a gaseous fluid source
for Einstein's equations.

In particular, we assume that the adiabatic index $\gamma$ may take particular values, excluding regions in which it cannot span.
To this end, we try to give either a thermodynamic explanation for Dark Energy or to formulate a cosmological model from basic
principles which take into account the laws of thermodynamics. We explore both the cases of varying and constant adiabatic indices and
find cosmological models which differ slightly from the concordance paradigm. In this way we provide a new approach in which dark
energy \emph{emerges} as a consequence of universe's thermodynamics. We propose tight bands of available values for the
adiabatic index and describe how to determine the difference between our thermodynamic Dark Energy contribution from a pure
cosmological constant, even at the level of background cosmology.

We investigate either the late-time or early-time universe and we show that our model is compatible with the basic requirements of the
standard paradigm. We noticed that our approach becomes a pure dark fluid contribution as the adiabatic index runs to vanish.  Finally,
we compare our approach with data, by means of SNIa, $H(z)$ and baryonic acoustic oscillation (BAO) data sets. Our numerical results
are compatible with the standard model, showing that our paradigm works fairly well in describing the universe expansion history at
different stages of universe's evolution. Slight departures are accounted in the shift of linear perturbations, whose corresponding
plots are inside the $10\%$ of discrepancies with respect to the $\Lambda$CDM model.

The paper is structured as follows. In Sec. (\ref{sec:heat_capacities}), we consider the properties of heat capacities in the context
of a homogeneous and isotropic universe. We describe how to build up physical definitions for them and how to obtain the corresponding
adiabatic indices, emphasising how to understand their physical meaning. In Sec. (\ref{sec:therm_adiab}), we describe the cosmological
consequences of inducing Dark Energy by examining either the case of constant or variable indices. We discuss the case of a purely
gaseous Dark Energy contribution and how to obtain the liming case of a dark fluid as $\gamma\rightarrow0$. In Sec.
(\ref{sec:early_universe}), we investigate some consequences of our approach in the high redshift regime, while in Sec.
(\ref{sec:constraints}), we consider two fitting procedures involving supernova and BAO data in order to obtain observational
constraints for our model. Finally, in Sec. (\ref{sec:conclusions}), we present our conclusions and give some perspectives for future
work.

\section{Heat capacities in observable cosmology}
\label{sec:heat_capacities}

In this section, we apply standard thermodynamics to the case of a homogeneous and isotropic universe in order to obtain expressions
for the specific heats. The thermodynamic laws can be used either in a classical or quantum regime, assuming that the universe does not
allow for the exchange of heat with the environment. It follows that the simplest choice for modelling the universe takes
into account that its evolution is purely adiabatic. While matter creation may occur within this framework \cite{alca}, we do
not assume this possibility in our approach.

In the case of a pure FLRW line element
\begin{equation}\label{frwmetric}
ds^2=dt^2-a(t)^2\left[\frac{dr^2}{1-k\,r^2}+r^2\left(\sin^2\theta d\phi^2+d\theta^2\right)\right]\,,
\end{equation}
the dynamics of the universe obeys the Friedmann equations:
\begin{subequations}\label{equazioni}
\begin{align}
H^2 &= \frac{\dot a^2}{a^2} = \frac{8\pi G}{3}\rho + \frac{\Omega_k H_0^2}{a^2}\,,\label{equazionenumero1}\\
\dot H + H^2 &= \frac{\ddot a}{a} = -\frac{4\pi G}{3} \left(\rho + 3P\right) \,,\label{equazionenumero2}
\end{align}
\end{subequations}
where dots represent derivatives with respect to the cosmic time $t$.

From Eq. \eqref{equazionenumero2}, it is easy to show that the present day dynamics of the universe is sourced by a perfect fluid whose equation of state is negative, to guarantee that $\ddot a>0$.

In general, the gravitational field is determined by the same source at all stages of its evolution, so that in order to guarantee
that at both early and late times the source is the same perfect fluid, one is forced to assume that it behaves like a thermodynamic
system \cite{bll}, whose evolution is described in terms of the redshift. Hence, we expect the laws of thermodynamics to hold
and to be mathematically consistent with the FLRW universe \cite{herny}.

However, the problem with the standard requirements imposed by thermodynamics in a FLRW universe is that it is difficult to construct
a self-consistent definition of temperature, because eventual departures from the background radiation temperature must be
associated with cosmological fluids \cite{boul}. For example, it has been recently proposed that the specific heats of the universe, given by:
\begin{subequations}
\begin{align}
C_V = \frac{\partial U}{\partial T}\,,\label{cv}\\
C_P = \frac{\partial h}{\partial T}\,,\label{cp}
\end{align}
\end{subequations}
are compatible with FLRW cosmology, but $T$ needs to be fixed to a precise value inside viable intervals. In general,
a possible Dark Energy temperature may evolve as the universe expands, so that $T$ could be considered as a function of the redshift
rather than a constant. For these reasons, a direct comparison of Eqs. \eqref{cv} - \eqref{cp} may be affected by theoretical
shortcomings on its validity.

In principle, both the internal energy and enthalpy are functions of the volume, pressure and temperature.
In this case, one naturally obtains:
\begin{subequations}\label{calorrr2}
\begin{align}
C_V &= \frac{1}{T'}\left[h' - \left(\frac{\partial U}{\partial T}\right)_V\left(\frac{\partial P}{\partial S}\right)_V V'-VP'\right]\,,\\
C_P &= C_V + \frac{1}{T'}\left[\left(\frac{\partial U}{\partial T}\right)_V\left(\frac{\partial P}{\partial S}\right)_V V'+VP'-\left(\frac{\partial h}{\partial P}\right)_T P'\right]\,.
\end{align}
\end{subequations}
However, since all state variables evolve in terms of the redshift, it seems natural to assume the simplest hypothesis in which both
energy and enthalpy are functions of $T$ only. To do so, it is possible to assume $U=\rho V$ and $h=(\rho+P)V$, where $V$ is the
volume of the universe. In this way, one splits the functional dependence of $U,h$ in terms of $V$, assuming that $\rho=\rho(T)$
and $P=P(T)$. The standard definition of the volume is $V=V_0a^{3}$ \cite{young}, which represents the simplest assumption
reflecting both early and late times of universe's evolution. Even though its introduction seems natural, other approaches
suggest alternative forms of the volume, for example an apparent horizon volume definition $V\propto r^{3}=V_0\,H^{-3}$
would reproduce a causal region where the entropy becomes $\propto H^{-2}$ \cite{horizon}. Moreover, employing the weak energy
conditions
\begin{eqnarray}\label{ECs0}
&T_{\mu\nu}k^\mu k^\nu\geq0 \,,\\
&\rho\geq 0\,, \\
&\rho + P \geq 0 \,,\\
\end{eqnarray}
one finds that both $U$ and $h$ must be positive definite.

Under the simplest hypothesis, in which Dark Energy weakly interacts, state functions depend on the temperature only and $V\propto a^3$, one simply obtains
\begin{subequations}\label{definitions}
\begin{align}
C_V &=\frac{3V_0}{8\pi G T'(1+z)^4}\left[2(1+z)HH'-3H^2 + \Omega_k(1+z)^2\right]\,,\\
C_P &= C_V + \frac{V_0}{T'(1+z)^3}\left(P'-\frac{3P}{1+z}\right)\,,
\end{align}
\end{subequations}
with no restrictions on their evolutions for different epochs of the evolution of the universe. Here primes denote derivatives with
respect to redshift $z$. The above forms of the heat capacities are not easy compare to cosmic predictions, due to the complexity
of their dependence on $H$ and its derivatives. One intriguing way to investigate their physical consequences is to frame
$C_V$ and $C_P$ in terms of observable quantities. In order to achieve this, let us consider expanding the scale factor $a(t)$
\cite{cosmo1}:
\begin{equation}
a(t)   \sim  1+  H_0 \Delta t - \frac{q_0}{2}  H_0^2\Delta t^2+\frac{j_0}{6} H_0^3 \Delta t^3 +  \ldots\,,
\end{equation}
where
\begin{subequations}\label{CSdef}
\begin{align}
H &= \frac{1}{a} \frac{d a}{d t}\,,\\
q &= -\frac{1}{a H^2} \frac{d^2a}{d t^2}\,,\\
j &= \frac{1}{a H^3} \frac{d^3a}{d t^3}\,,
\end{align}
\end{subequations}
and since \cite{cosmo4}
\begin{equation}\label{acca1}
H= H_0\left(1+\left.\frac{H'}{H_0}\right|_{z=0}\,z + \left.\frac{H''}{2H_0}\right|_{z=0}\,z^2+\dots \right)\,,
\end{equation}
by comparing Eqs. \eqref{CSdef} with Eq. \eqref{acca1}, we get \cite{cosmo5}:
\begin{equation}\label{esx}
H' \equiv \frac{1+q}{1+z}\,H\,,\quad H'' \equiv \frac{j-q^2}{(1+z)^2}\,H\,,
\end{equation}
and by virtue of Eqs. \eqref{definitions}, we obtain:
\begin{subequations}
\begin{align}
C_P&=\frac{2V_0}{T'}\frac{(j-1)H^2 +\Omega_k H_0^2 (1+z)^2}{(1+z)^4}\,,\label{calorecosmografico1}\\
C_V&=\frac{3V_0}{T'}\frac{(2q-1)H^2 + \Omega_k H_0^2(1+z)^2}{(1+z)^4}\,.\label{calorecosmografico2}
\end{align}
\end{subequations}
The above expressions give us direct information on the form of Dark Energy in cases where one is able to describe the
temperature as a function of redshift $z$.

This treatment is essentially based on the standard requirements for obtaining an evolving adiabatic index from
the ratio between heat capacities.

Using~(\ref{calorecosmografico1}) and~(\ref{calorecosmografico2}), the adiabatic index becomes:
\begin{equation}\label{adiabaticindex}
\gamma = \frac{2\left[(j-1)H^2 + \Omega_k H_0^2(1+z)^2\right]}{3\left[(2q-1)H^2+ \Omega_k H_0^2(1+z)^2\right]}\,.
\end{equation}
In the simplest case of $\Omega_k=0$, the three allowed regimes are: $0<\gamma<1$, $\gamma=1$ and $\gamma>1$.
The consequences on the thermodynamics of Dark Energy are summarised as follows: $C_V,C_P<0$ in the first case, $C_P=0$ and
$C_V<0$ in the second case and $C_V,C_P>0$ in the last case.

When $\Omega_k=0$, there also esists a region for which $C_P=0$ and $C_V=0$.  This occurs when $q\to 1/2$ and $j\to 1$. In general, this could happen at a redshift $z\gg1$, under the hypothesis of de-Sitter contribution to Dark Energy.

Another interesting fact is how $\gamma$ behaves at the transition redshift $z_\text{tr}$ \cite{transition}, i.e.,
when Dark Energy starts to dominate over matter. Assuming for simplicity that $\Omega_k=0$, the adiabatic index becomes:
\begin{equation}\label{adiabatictransition}
\gamma_{\rm tr}=-\left.\frac{8\pi G(1+z_{\rm tr})P'}{3H^2}\right|_{z=z_{\rm tr}} = \left.(1+z_{\rm tr})\frac{d\ln P}{dz}\right|_{z=z_{\rm tr}}\,,
\end{equation}
where we have made use of $P=8\pi G H^2(2q-1)$ which is a direct consequence of Eq. (\ref{equazionenumero2}), and $q(z_{\rm tr})=0$. From the above
considerations, it is easy to see that the expression \eqref{adiabatictransition} is valid for \emph{any} cosmological model.
This heuristically shows that the adiabatic index is intimately related to the variation of the pressure. As a consequence,
via the dynamics of the Friedmann equations, this determines how Dark Energy evolves and how possible departures from the standard
concordance model may arise. This issue will be addressed in the next sections, in which we investigate the consequences of Eq.
 \eqref{adiabaticindex} in cosmology.

\section{Thermodynamics of adiabatic Dark Energy}
\label{sec:therm_adiab}

Standard thermodynamics states that the combination of the first and second principles leads to
\begin{equation}
TdS=d(\rho V)+P dV = d[(\rho+P)V]-V dP\,,
\end{equation}
and since $\frac{\partial^2 S}{\partial T \partial V}=\frac{\partial^2 S}{\partial V \partial T}$, one gets $dP=(\rho + P)dT/T$.
It is therefore easy to demonstrate that:
\begin{equation}\label{entropia}
dS=d\left[\frac{(\rho+P)V}{T}\right]\,,
\end{equation}
where any arbitrary constant is assumed to be zero for simplicity. The basic requirements of thermodynamics suggest that
$S \equiv V(\rho + P)/T$. Taking the combination of the first and second Friedmann equations, one gets:
\begin{equation}\label{continuity}
\dot \rho+3H(P+\rho)=0\,,
\end{equation}
which can be recast as $d(\rho V) + P dV = 0$.

The conservation law, by virtue of Eq.~(\ref{entropia}) becomes
\begin{equation}
d\left[\frac{(\rho+P)V}{T}\right]=0,
\end{equation}
leading to the fact that $S=$ const. This is equivalent to a thermodynamic system in which there is no heat exchange, i.e., is
adiabatic

The complete system, rewritten in terms of the redshift, is
\begin{subequations}
 \begin{align}
 & \frac{H^2}{H_0^2} = \frac{8\pi G\rho}{3H_0^2} + \Omega_m^{(\text{ext})}(1+z)^3 + \Omega_k(1+z)^2\,, \label{fried_1_z}\\
 & (1+z)HH' - H^2 = 4\pi G\left[P + \frac{\rho}{3} + \Omega_m^{(\text{ext})} H_0^2(1+z)^2\right]\,, \label{fried_2_z}\\
 & P = P_0 V^{-\gamma} = P_0(1+z)^{3\gamma}\,, \label{P_V_gamma}\\
 & \gamma = \frac{C_P}{C_V} = \frac{(\rho'+P')V + (\rho + P)V'}{\rho' V + \rho V'}\,,\label{gamma_CP_CV}
\end{align}
\end{subequations}
where we have allowed for an \emph{external} non-relativistic matter component $\Omega_m^{(\text{ext})}$, which in principle can be chosen arbitrarily in a way that will become clearer later.

From~(\ref{fried_1_z}) and (\ref{fried_2_z}) we can simplify~(\ref{gamma_CP_CV}) to obtain:
\be
\gamma = \frac{(\rho'+P')V + (\rho + P)V'}{\rho' V + \rho V'} = -\frac{P' V}{P V'} = \frac{(1+z)P'}{3P}\,.\label{gamma_cond_1}
\ee
On the other hand, taking~(\ref{P_V_gamma}) gives
\be
\frac{(1+z)P'}{3P} = \gamma + (1+z)\gamma'\ln(1+z)\,.
\ee
Clearly, this is compatible with~(\ref{gamma_cond_1}) only for $\gamma' = 0$, that is $\gamma = $ constant. In this case,
it is easy to see that $\rho$ is given by
\be\label{eq_rho_of_z}
\rho(z) = \left(\rho_0 + \frac{P_0}{1-\gamma}\right)(1+z)^3 - \frac{P_0}{1-\gamma}(1+z)^{3\gamma}\,.
\ee
First of all, we notice that there appears a term scaling as $(1+z)^3$, which corresponds to non-relativistic matter; this
term can in principle be identified with cosmological dark matter, but not necessarily so (see below). Moreover, there appears a second term which instead scales as $(1+z)^{3\gamma}$. Choosing $\gamma = 0$ one recovers the dark fluid~\cite{ref:dark_fluid}, whereby the corresponding term assumes the role of a pure $\Lambda$ term.

Solutions with $\gamma$ evolving with redshift are in principle possible, although in order for this to be possible one has
to relax at least one condition between~(\ref{P_V_gamma}) and~(\ref{gamma_CP_CV}). For instance, one could try defining the
polytropic behaviour using $P\sim \rho^\gamma$ instead of $P\sim V^{-\gamma}$. These ideas will be dealt with in future works.

From~(\ref{P_V_gamma}) and~(\ref{eq_rho_of_z}) we find that the equation of state $w\equiv P/\rho$ of the adiabatic fluid evolves as
\be
w(z) \equiv \frac{P(z)}{\rho(z)} = -\frac{(1-\gamma)w_0(1+z)^{3\gamma}}{w_0(1+z)^{3\gamma}-(1-\gamma + w_0)(1+z)^3}\,.
\ee

Defining
\be\label{eq:Omega_m}
\Omega_m \equiv \frac{8\pi G}{3H_0^2}\left(\rho_0 + \frac{P_0}{1-\gamma}\right) + \Omega_m^{(\text{ext})}\,,
\ee
we can write the Hubble parameter in the simple form
\be\label{eq:Hubble_exact}
\frac{H(z)^2}{H_0^2} = \Omega_m(1+z)^3  + \Omega_k(1+z)^2 + (1-\Omega_m - \Omega_k)(1+z)^{3\gamma}\,.
\ee
The analysis presented in section~\ref{sec:constraints} will be performed using $\Omega_m$, $\Omega_k$ and $\gamma$ as the
independent parameters. The parameter $\Omega_m$ describes the total ``matter'' content of the universe, be it external
(baryons and/or Dark Matter) or due to the evolution of the polytropic fluid under study. As one can see from~(\ref{eq:Omega_m}), fixing $\Omega_m$ and $\gamma$ still leaves freedom in choosing the value of $\Omega_m^{(\text{ext})}$ and $w_0 = P_0/\rho_0$,
so essentially one can insert the desired amount of external non-relativistic matter by hand. This is a very
interesting result because choosing any value of $\Omega_m^{(\text{ext})}$, the ``right'' amount of dust-like fluid can automatically be accounted for. The most relevant possibilities are:
\begin{itemize}
 \item $\Omega_m^{\rm (ext)} = \Omega_m$, that is
 \be
 w(z) = w_0 = \gamma - 1\,.
 \ee
 We are basically tuning our fluid so that its dust component vanishes. In this picture, the new fluid only contributes to Dark Energy, and baryons and CDM are assumed as external and independent components. Moreover, the Dark Energy component has constant equation of state $w = \gamma - 1$. Notably, in this case our model is equivalent to $\omega$CDM.

 \item $\Omega_m^{\rm (ext)} = \Omega_b$: in this case, the Universe is filled with just baryons and the new fluid, which is mimicking both Dark Energy and cold Dark Matter. For fixed $\gamma$, $\Omega_m$ and $\Omega_b$, we must choose $P_0$ so that
 \be
 \frac{8\pi G}{3H_0^2}\left(\rho_0 + \frac{P_0}{1-\gamma}\right) = \Omega_m - \Omega_b \equiv \Omega_{\rm CDM}\,,
 \ee
 which corresponds to
 \be
 P_0 = (\gamma - 1)\left(\rho_0 - \frac{3H_0^2}{8\pi G}\,\Omega_{\rm CDM}\right)\,.
 \ee
\end{itemize}
Let us stress that these two possibilities, and indeed any other combination of $P_0$ and $\Omega_m^{\rm (ext)}$ resulting in the same $\Omega_m$, will not need separate analyses. As shown in~(\ref{eq:Hubble_exact}), $\Omega_m$ and $\gamma$ are the only parameters associated to the fluid relevant for cosmological fits.

\section{Cosmological constraints}
\label{sec:constraints}

We test our model using a Metropolis-Hastings Monte Carlo code using $L = \exp(-\chi^2_\text{tot}/2)$ as likelihood function,
with
\[
\chi^2_\text{tot} = \chi^2_\text{SN} + \chi^2_{H} + \chi^2_\text{BAO}\,,
\]
equipped with a Gelman-Rubin convergence diagnostic. We use several cosmological data sets: SNIa data from the Union2.1
compilation~\cite{ref:union2.1}, $H(z)$ data (as quoted in~\cite{ref:Luongo_H(z)}), and
BAO~\cite{Beutler:2011hx,Anderson:2013zyy,Blake:2011en}.

\subsection{Differential age and \texorpdfstring{$H(z)$}{H} data}

Independent reconstructions of Hubble measurements constitute a novel approach to track the evolution of the universe
without invoking a model \emph{a priori}. In particular, employing massive early-type galaxies as cosmic chronometers,
it would be possible to match astronomical and cosmological measurements to evaluate the differential age, i.e., the
ratio $\frac{dt}{dz}$. Since, differentiating the scale factor definition with respect to the redshift, one obtains
\begin{equation}\label{HHH}
\frac{dz}{dt}=-(1+z)H(z)\,,
\end{equation}
it is possible, knowing the redshift at which the measure has been performed, to evaluate $H$ at different stages of the
universe's evolution.

We compare $H(z)$ data with the exact solution~(\ref{eq:Hubble_exact}). The $\chi^2$ is
\be\label{eq:chi^2_H(z)}
\chi^2_H = \sum_i \frac{\left[H^\text{(model)}(z_i) - H^\text{(exp)}(z_i)\right]^2}{\sigma_i^2}\,.
\ee

\subsection{Supernovae Ia data}

We use SNIa data from the Union2.1 catalogue, containing 580 data points. Type Ia supernovae observations have been
extensively employed during the last few decades for cosmological model parameter-fitting. Supernovae Ia are widely
thought to be standard candles, i.e., objects whose luminosity curves are intimately related to distances\footnote{However, it
should be stressed that SNIa absolute magnitudes can be neither directly measured nor inferred from theoretical considerations
with arbitrary accuracy. See also below.}. The Union2.1 catalogue is built up to extend previous versions, with the advantage
that the whole systematics is mostly reduced. So that, one can assume that systematic errors do not influence numerical outcomes.
Moreover, errors on the redshift measurements are assumed to be negligibly small.

The observable quantity associated to SNIa is the distance modulus, namely the difference between the apparent magnitude $m$
and absolute magnitude $M$ of each object:
\be
\label{eq:distance_modulus}
\mu(z) \equiv m(z) - M = 5\log_{10}\frac{d_L(z)}{10\,\text{pc}}
\ee
where
\be
d_L(z) = \frac{(1+z)}{H_0}\times
\begin{cases}
 \cfrac{1}{\sqrt{\Omega_k}}\sin\left[\sqrt \Omega_k\,D(z)\right] & \Omega_k > 0\\
 D(z) & \Omega_k = 0 \\
 \cfrac{1}{\sqrt{|\Omega_k}|}\sinh\left[\sqrt{|\Omega_k|}\,D(z)\right] & \Omega_k < 0
\end{cases}
\ee
and
\be
D(z) = H_0\int_0^z \frac{1}{H(z')}\,.
\ee
However, since $M$ is not known with sufficient accuracy from theoretical arguments, Union2.1 data are only
reliable up to a normalisation:
\be\label{eq:nuisance}
\mu_\text{obs}(z;\mathcal M) = \mu_\text{Union2.1}(z) + \mathcal M\,,
\ee
where $\mu_\text{Union2.1}$ is the value reported in the Union2.1 compilation, and $\mathcal M$ depends both on the
absolute magnitude of supernovae Ia and on $H_0$, but it does not affect the expansion history $H(z)/H_0$. It will have to be
treated as a \textit{nuisance parameter}, fitted along with the other cosmological parameters and marginalised over.

The distance modulus for a given redshift and set of parameters is computed via numerical integration
using~(\ref{eq:Hubble_exact}). The total $\chi^2_\text{SN}$ is computed analogously to~(\ref{eq:chi^2_H(z)}):
\be
\chi^2 = \sum_i \frac{[\mu_\text{th}(z_i) - \mu_\text{obs}(z_i;\mathcal M)]^2}{\sigma_i^2}\,.
\ee

\subsection{Baryonic Acoustic Oscillations (BAO)}

Baryonic acoustic oscillations (BAO) are observed in large scale structure (LSS), and are the result of sound waves propagating in the early universe. In recent years, they have provided us with another relevant data set for cosmological fits. Measuring the position of the BAO peak in the LSS correlation function corresponds to measuring a combination of angular distance and luminosity distance, namely
\be\label{eq:DV}
D_V^3(z) \equiv \frac{d_L^2(z)\,z}{(1+z)^2H(z)}\,.
\ee
This quantity tracks the comoving volume variation at a given redshift.

We will consider the two BAO observables
\be
A(z) \equiv \frac{H_0 D_V(z) \sqrt{\Omega_m}}{z}\,,\quad d_z(z) \equiv \frac{r_s(z_\text{drag})}{D_V(z)}\,,
\ee
where $r_s(z_\text{drag})$ is the comoving sound horizon at the baryon drag epoch. This quantity needs to be calibrated with CMB data assuming a fiducial cosmological model, with the latest data giving
\be\label{eq:sound_horizon_drag}
\begin{aligned}
 & z_\text{drag} =
 \begin{cases}
  1020.7 \pm 1.1 & \text{WMAP9~\cite{Bennett:2012zja}}\\
  1059.62 \pm 0.31 & \text{Planck~\cite{Ade:2015xua}}
 \end{cases}\\
 & r_s(z_\text{drag}) =
 \begin{cases}
  152.3 \pm 1.3 & \text{\cite{Bennett:2012zja}}\\
  147.41 \pm 0.30 & \text{\cite{Ade:2015xua}}
 \end{cases}
\end{aligned}
\ee
In this sense, BAO data are slightly model-dependent, since acoustic scales depend on the redshift (drag time redshift), inferred from first order perturbation theory assuming a given cosmology. However, the same data would still be reliable when studying any realistic cosmology which differs from $\Lambda$CDM only at relatively low redshifts.
For the fits, we will thus use a gaussian prior using the Planck best value.

As for the case of SN data, the theoretical values of $A(z)$ of $d_z$ are computed via numerical integration using the exact expression for the Hubble parameter~(\ref{eq:Hubble_exact}). We use the BAO data shown in table~\ref{tab:BAO}.
\begin{table}[tb]
\begin{tabular}{c c c c c}
\hline
Experiment & $z$ & $d_z \pm \sigma_d$ & $A(z) \pm \sigma_A$ & Ref.\\
\hline
6dFGS & 0.106 & $0.336\pm 0.015$ & & \cite{Beutler:2011hx}\\
SDSS-III & 0.32 & $0.1181 \pm 0.0023$ & &\cite{Anderson:2013zyy}\\
 & 0.57 & $0.0726 \pm 0.0007$ & &  \\
WiggleZ & 0.44 & & $0.474 \pm 0.034$ & \cite{Blake:2011en}\\
& 0.6 & & $0.442 \pm 0.020$ & \\
& 0.73 & & $0.424 \pm 0.021$ & \\
\hline
\end{tabular}
\caption{BAO data used in the analysis. For each experiment, we quote the observable more suitable for the analysis.}
\label{tab:BAO}
\end{table}
Not all data are uncorrelated. In fact, the covariance matrix (symmetric, we only quote the upper diagonal) for the WiggleZ data at $z=(0.44,0.6,0.73)$ is~\cite{Blake:2011en}:
\be
C^{-1} =
\begin{pmatrix}
 1040.3 & -807.5 & 336.8\\
  & 3720.3 & -1551.9\\
  & & 2914.9
\end{pmatrix}
\,.
\ee
The total $\chi^2$ for BAO data is
\be
\chi^2_\text{BAO} = \chi^2_\text{6dFGS} + \chi^2_\text{SDSS-III} + \chi^2_\text{WiggleZ}\,,
\ee
with
\be
\begin{aligned}
& \chi^2_\text{BAO,SDSS-III} = \sum \left[\frac{d_z^\text{obs} - d_z^\text{th}(z_i)}{\sigma_d}\right]^2\,,\\
& \chi^2_\text{WiggleZ} = (\mathbf A^\text{obs} - \mathbf A^\text{th})^T C^{-1} (\mathbf A^\text{obs} - \mathbf A^\text{th})
\end{aligned}
\ee

\subsection{Results}

\begin{figure}[t]
\includegraphics[scale=.7]{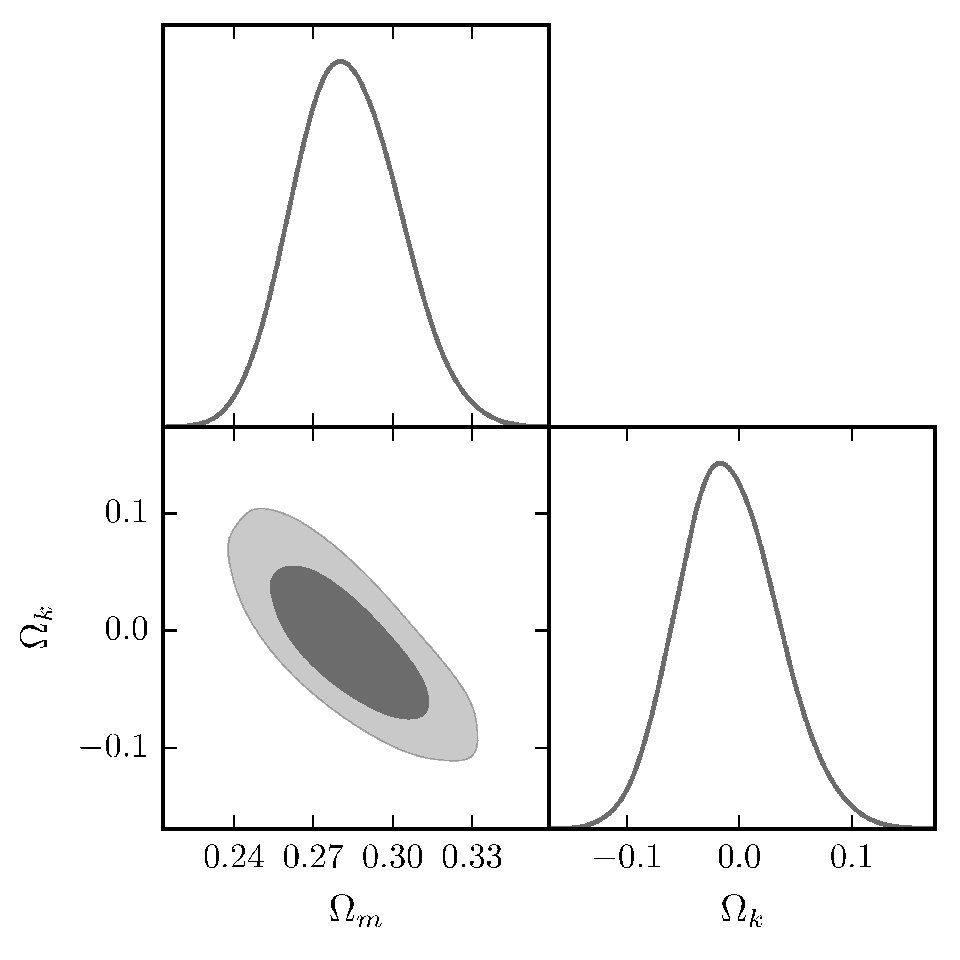}
\hspace*{.5em}
\includegraphics[scale=.7]{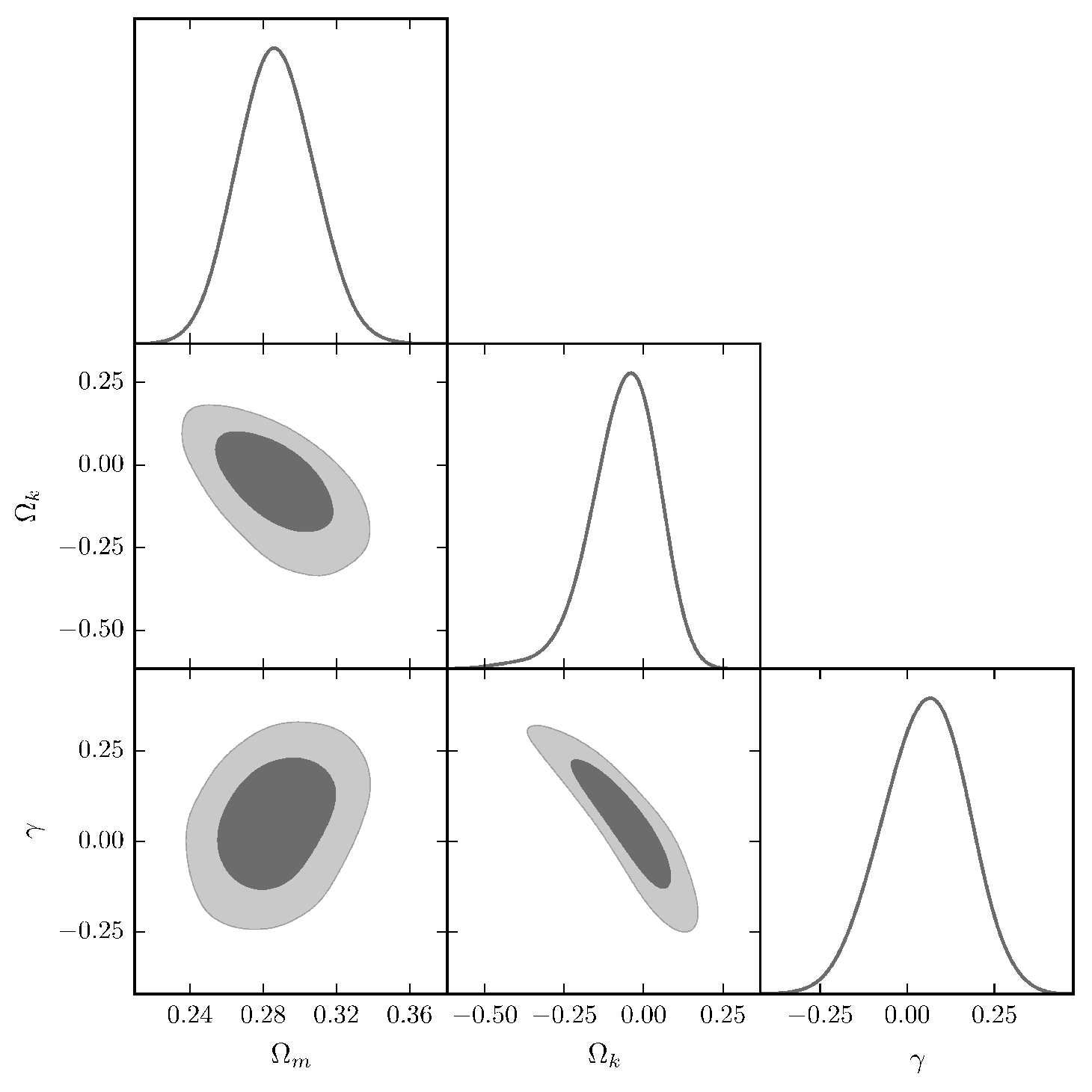}
\caption{$1\sigma$ and $2\sigma$ contours corresponding to $\gamma = 0$ ($\Lambda$CDM) (left), and to our model with
$\gamma$ free (right). See also Tab.~\ref{tab:results}.}
\label{fig:results}
\end{figure}

Here we use flat priors on the fitting parameters, gaussian priors on $r_s(z_\text{drag})$ as mentioned above, and we marginalise
over the nuisance parameter $\mathcal M$. We show results for the broad prior $\Omega_k ={\rm Uniform}(-1,1)$ and $\Omega_k = 0$ in figure~\ref{fig:results}. The corresponding means, 95\% confidence levels and best fits are shown in table~\ref{tab:results}.
\begin{table}[b]
\begin{tabular}{c c}
\begin{tabular}{c c c c}
Parameter & Prior & 95\% Limits & Best fit\\
\hline
$\Omega_m$          & Uniform(0,1)  & $0.283^{+0.040}_{-0.037}$  & 0.2843 \\
$\Omega_k$          & Uniform(-1,1) & $-0.009^{+0.090}_{-0.082}$ & -0.0174\\
$\gamma$            & $= 0$         &                            &  \\
$\chi^2_\text{tot}$ &               &                            & 593.270 \\
\hline
 \end{tabular}

&
\begin{tabular} { c c c c}
Parameter & Prior & 95\% Limits & Best fit\\
\hline
$\Omega_m$          & Uniform(0,1) & $0.287^{+0.041}_{-0.039}$ & 0.2865 \\
$\Omega_k$          & Uniform(-1,1) & $-0.06^{+0.20}_{-0.22}$   & -0.05483 \\
$\gamma$            & Uniform(-1,1) & $ 0.05^{+0.23}_{-0.24}$   & 0.05877 \\
$\chi^2_\text{tot}$ &              &                           & 592.984 \\
\hline
 \end{tabular}
\\
$\Lambda$CDM ($\gamma = 0$) & $\gamma \neq 0$
\end{tabular}
\caption{Results for our model for $\Lambda$CDM, i.e. $\gamma = 0$ (left), and $\gamma$ generic (right). See
also Figure~\ref{fig:results}.}
\label{tab:results}
\end{table}
We can see that results are compatible with $\gamma = 0$, which corresponds to the $\Lambda$CDM or dark
fluid~\cite{ref:dark_fluid} solution, and with $\Omega_k = 0$ which is assumed in many cosmological analyses.
As expected, $\gamma$ is constrained to small values, roughly $|\gamma|\lesssim 0.25$ at $2\sigma$.

Considering only the $\chi^2$ value, our modified cosmology appears to be only slightly preferred over the standard one, as one
can see comparing the left and right tables in Tab.~\ref{tab:results}. These considerations suggest that it would be useful to consider a a discussion on model selection criteria for our approach and the concordance model. See the next section for details.

\section{Model selection criteria}
\label{sec:bayes}

For much of the community, the $\Lambda$CDM paradigm is the favourite framework to fit cosmic data, due to its simplicity
and the fact that the only free parameters are $\Omega_m$ and $\Omega_k$. However, a large number of different possibilities
go beyond this choice and
therefore one needs to determine methods able to compare the range of different competing cosmological models. Two statistical
model independent methods are offered by the so called selection criteria, aimed at determining the "best" model by considering
the combination of $\chi^2$ and degrees of freedom. This turns out to be important, since it is possible that
viable models with higher orders of parameters provide higher chi squares as well, which do not have to be considered
as disfavoured at all than the standard model.

\noindent Three suitable selection criteria are the Akaike information criterion (AIC), the corrected AIC (AICc) and the Bayesian
information criterion (BIC)~\cite{AIC}. These tests are a standard diagnostic tool \cite{do} of regression
models \cite{tro}. They are defined as
\begin{subequations}
 \label{eq:AIC_BIC}
 \begin{align}
  & \text{AIC}  \equiv -2\ln \mathcal L + 2d\;,\\
  & \text{AICc} \equiv \text{AIC} + \frac{2d(d+1)}{N-d-1}\;,\\
  & \text{BIC}  \equiv -2\ln \mathcal L + 2d\ln N\;,
 \end{align}
\end{subequations}
where
\be
\mathcal L = \exp(-\chi^2/2)
\ee
is the chosen likelihood function, $d$ is the number of model parameters and $N$ is the number of data points, which in our case is
\be
N = N_\text{SN} + N_H + N_\text{BAO} = 580 + 28 + 6 = 614\,.
\ee
The basic requirement is to essentially postulate two distribution functions, namely $f(x)$ and $g(x|\theta)$. Here, $f(x)$ is taken
to be the \emph{exact} distribution function, whereas $g(x|\theta)$ approximates the former. The way of approximating this makes use
of a set of parameters which has been denoted by $\theta$. Thus, there exists only a set
of $\theta_{min}$, which minimises the difference between $g(x,\theta)$ and $f(x)$ \cite{quqq}.

It follows that the AIC, AICc and BIC values for a single model are meaningless since the exact model function
$f(x)$ is unknown. For those reasons, one is only interested in the differences
\be
\Delta X = X_\gamma - X_{\Lambda\text{CDM}}\qquad\qquad X = \text{AIC, AICc, BIC}\;.
\ee
These quantities must be evaluated for the whole set of models involved in the analysis.

Results are shown in Tab.~\ref{tab:AIC_BIC}. The AIC(c) tests indicate a slight preference for $\Lambda$CDM, whereas
the BIC test suggests a very strong preference. Indeed, the BIC has a much stronger penalty for extra parameters, although
a pure Bayesian evidence analysis sometimes gives results more in line with the AIC(c).

\begin{table}[t]
\begin{tabular}{c c c c c c}
 Model          &  $\chi^2_\text{best fit}$  & $\Delta d$ &  $\Delta_\text{AIC}$  & $\Delta_\text{AICc}$ & $\Delta_\text{BIC}$ \\
 \hline
 $\Lambda$CDM   & 593.270     & 0   &  0 & 0 & 0\\
 $\gamma$       & 592.984    & 1   &  1.714 & 1.721 & 12.554\\
 \hline
\end{tabular}
\caption{Comparison between $\Lambda$CDM and our model using three criteria: AIC, AICc, and BIC. See text for details.}
\label{tab:AIC_BIC}
\end{table}

\section{Consequences on early-time cosmology}
\label{sec:early_universe}

Let us now investigate how the correction due to our Dark Energy model affects the universe dynamics at high redshifts.
To do so, we study the impact of the modified background evolution on density perturbations, which likely represent the most
suitable framework in which one can naively check the goodness of any cosmological model at high redshifts. The
perturbation equations, in their coarse-grained form, simply read
\begin{equation}\label{7}
\ddot{\delta}+2H\dot{\delta}-4\pi G\rho_{m}\delta=0\,.
\end{equation}
The so called \emph{growth evolution}, intimately related to $\delta$, may be easily handled by means of the scale
variable $\ln a$. One can parametrise the Dark Energy effects using the growth variable
\be\label{eq:growth_D}
D(a)=\frac{\delta}{a}
\ee
which satisfies:
\be
\label{final}
D''+D'\left[\frac{5}{a}+\frac{(\ln E^2)'}{2}\right] + \frac{D}{a}\left[ \frac{3}{a}\left(1-\frac{\Omega_m} {2E^{2}a^{3}} \right) + \frac{(\ln E^2)'}{2} \right]=0\,,
\ee
where a prime denotes differentiation with respect to the scale factor $a$, and $E\equiv H/H_0$. We assume the
boundary conditions $D(a_{LSS})=1$ and $D'(a_{LSS})=0$, with $a_{LSS}=(1+z_{LSS})^{-1}$, i.e., the last scattering surface
scale factor, approximated by $z_{LSS}\approx 1089$.

The growth history for a given $\gamma$ can be compared with the standard model, i.e., $\Lambda$CDM, keeping in mind that
any viable cosmological models should not yield deviations exceeding about $10\%$.

Analogously, one can define the growth index as
\be
f=\frac{d\ln \delta}{d\ln a}\,,
\ee
which enables one to rewrite the perturbation equations as:
\begin{equation}\label{growth index}
    f'+\frac{f^{2}}{a}+ \left[\frac{2}{a}+\frac{(\ln E^{2})'}{2}
    \right]f-\frac{3\Omega_{m}}{2E^{2}a^{4}}=0\,,
\end{equation}
with the boundary condition $f(a_{LSS})=1$. Furthermore, to corroborate the results on the shift parameters, we also
plot the relative deviation
\be\label{eq:deviation_H}
\Delta_H \equiv \frac{H_\Lambda − H_\text{model}}{H_\Lambda}
\ee
between the Hubble rates of the $\Lambda$CDM model ($\gamma = 0$) and that corresponding to a general $\gamma$.

\begin{figure}
\def\mywid{.31\textwidth}
\includegraphics[width=\mywid]{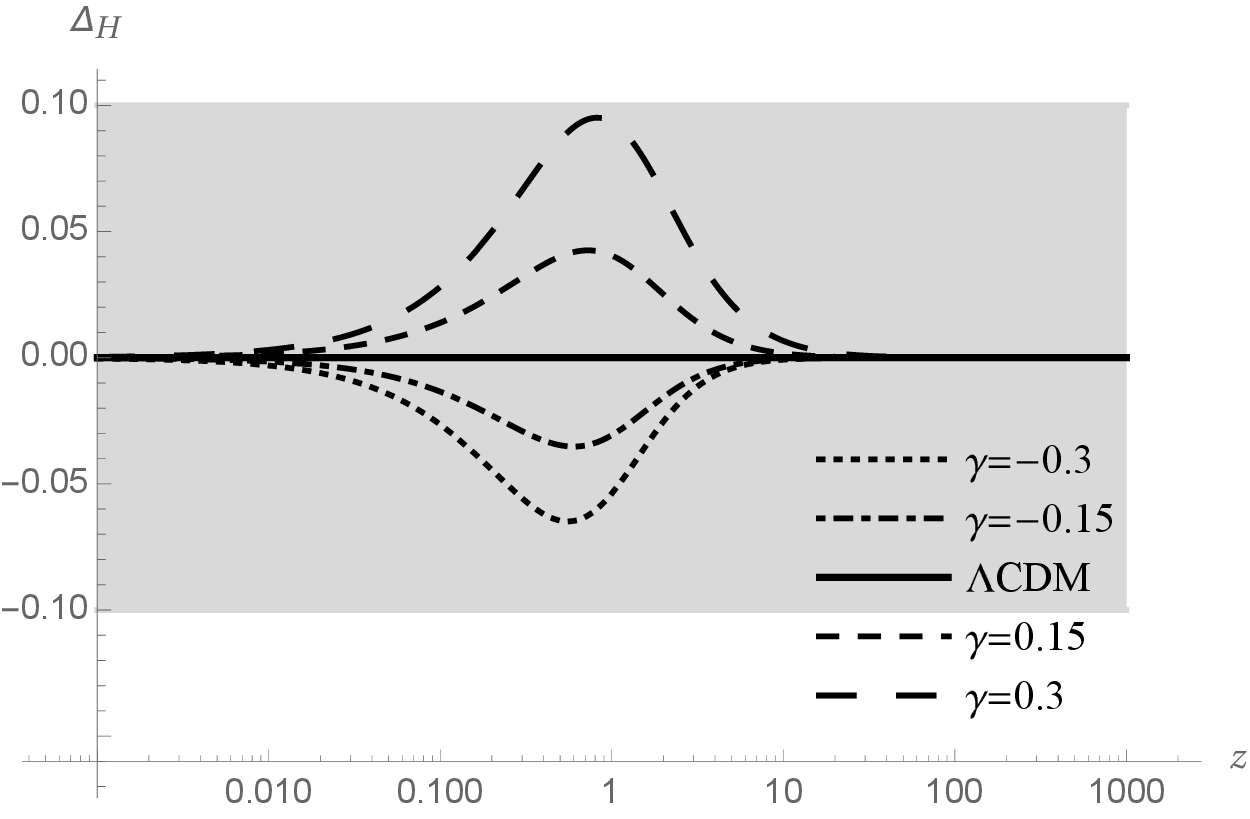}
\hspace*{1em}
\includegraphics[width=\mywid]{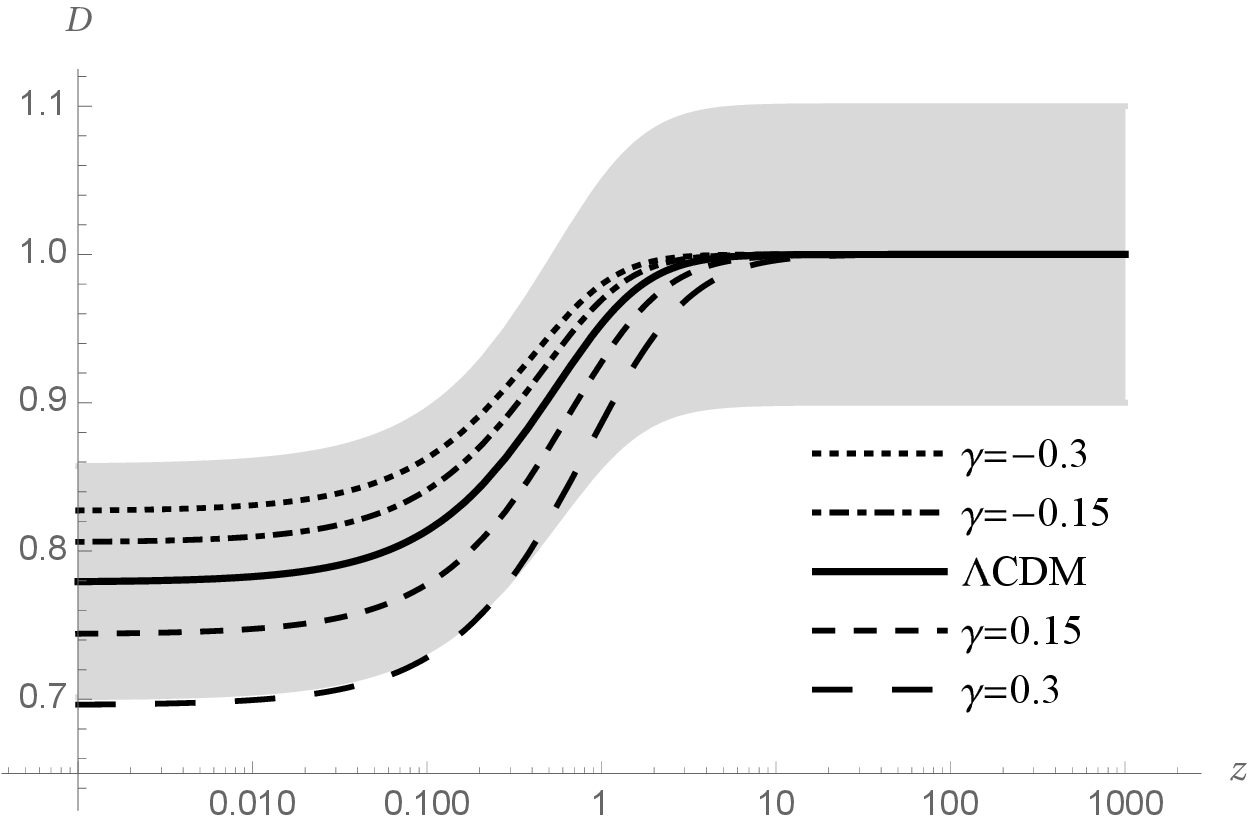}
\hspace*{1em}
\includegraphics[width=\mywid]{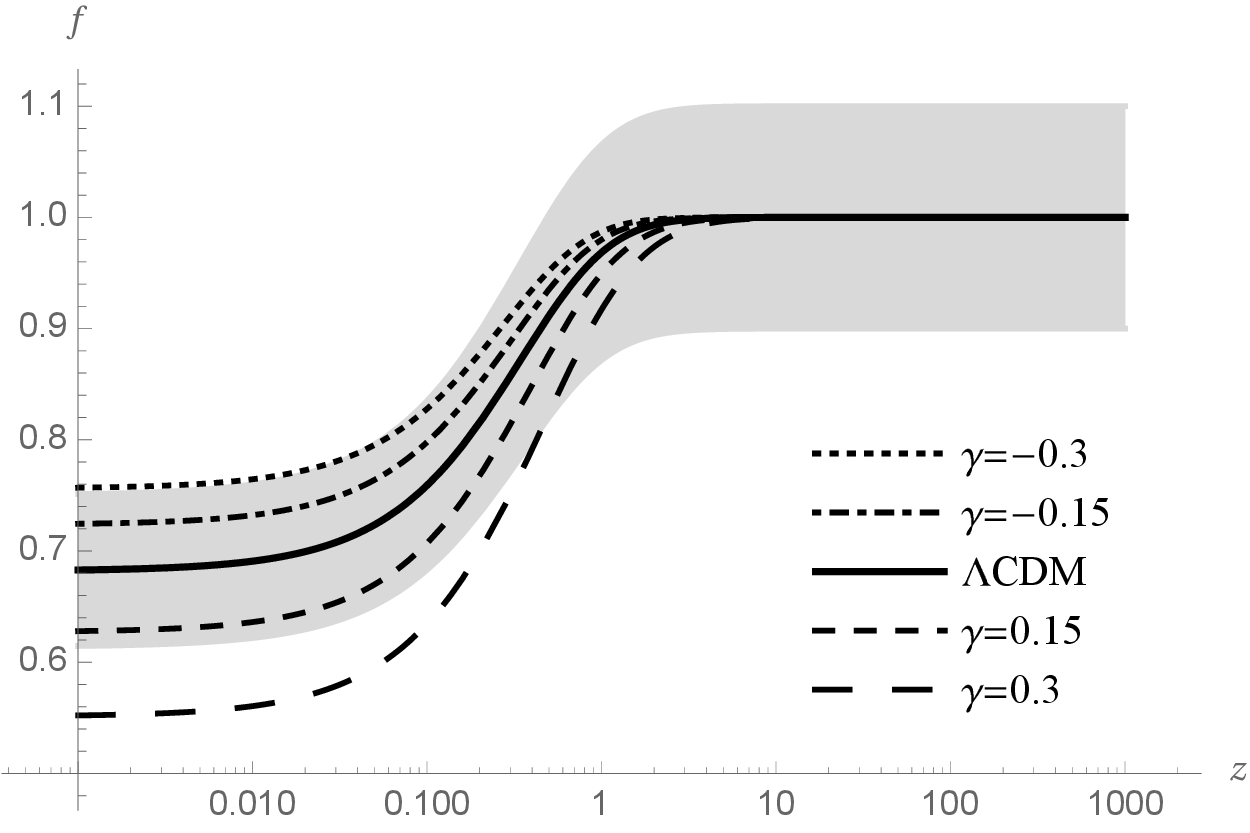}
\caption{Comparison between $\Lambda$CDM (black solid lines) and our model, for a few values of $\gamma$: -0.3 (dotted), -0.15 (dot-dashed), 0.15 (dashed, small) and 0.3 (dashed, large). The grey bands correspond to $\pm 10\%$ departures from $\Lambda$CDM. The plotted quantities are $\Delta_H$ (Eq.~\ref{eq:deviation_H}), $D$ (Eq.~\ref{eq:growth_D}), and $f$ (Eq.~\ref{growth index}). For definiteness, we have taken $\Omega_m = 0.3$ and $\Omega_k = 0$.}
\label{fig:smallpert}
\end{figure}

Plots of $D$, $f$ and $\Delta_H$ for a few values of $\gamma$, compared with the $\Lambda$CDM predictions are shown in Fig.~
\ref{fig:smallpert}. As we can see from numerical results, $\Delta_H$ differs substantially from zero only at intermediate
redshifts ($0.1\lesssim z \lesssim 10$), peaking around $z\sim 1$. At large redshifts, the Dark Energy component is
completely negligible, and our model is indistinguishable from $\Lambda$CDM. Naturally, we also have
$\Delta_H \to 0$ as $z\to 0$ because we require that $H(z\to 0) = H_0$ for any model.

$D$ and $f$ start departing from the standard $\Lambda$CDM solution around $z\sim 1$, after which they follow the standard
evolution but with a normalisation factor with respect to the standard cosmological scenario.

All curves for $\Delta_H$, $D$ and $f$ fit within the $\pm 10\%$ bands with the exception of the $\gamma = 0.3$ solution
in the plot for $f$. Notice that the 95\% limits from the numerical fits of Table~\ref{tab:results} constrain
$|\gamma|\lesssim 0.25$ so this result does not effectively reduce the allowed range of values for $\gamma$.

\section{Discussion and conclusions}
\label{sec:conclusions}

We have developed a simple Dark Energy model starting from an adiabatic fluid which evolves following rather standard
thermodynamic considerations. The adiabatic index $\gamma$ is expressed as a specific function of redshift and the Hubble parameter.
In a homogeneous and isotropic universe, solutions give $\gamma = $ constant, but we would expect this to change if one or more of the
assumptions are relaxed.

\noindent In the simplest case of $P\propto V^{-\gamma}$, the resulting fluid evolutions are characterised by the combination of two effective fluids evolving differently throughout the
history of the universe: a component scaling as $(1+z)^3$, i.e. as non relativistic matter, and another scaling as
$(1+z)^{3\gamma}$, which corresponds to a dynamical Dark Energy term. The $\Lambda$CDM paradigm is included in our model,
taking $\gamma = 0$. We found the freedom of choosing $P_0, \rho_0$ and $w_0$ according to the total amount of Dark Sector. In other words, using for example that the external matter is made by baryons only, we demonstrated that the Dark Matter amount is immediately recovered, without postulating it. In such a way, we showed that our model allows Dark Energy and Dark Matter to be emergent phenomena, described by standard laws of thermodynamics.

We fitted the model using SNIa, $H(z)$ and BAO data, and found the corresponding constraints on the model parameters.
As expected, the $\Lambda$CDM solutions fit perfectly within the observational bounds, and are actually preferred by model
selection criteria despite the slight improvement in terms of total $\chi^2$ for our model.

We also performed an analysis of the evolution of density perturbations at high redshifts and found that
reasonable values of $\gamma$ are well within the allowed discrepancies from the standard cosmological scenario derived from the late universe constraints.

In future works, it would be interesting to relax the hypothesis $P\propto V^{-\gamma}$ and explore the consequences on cosmology. Moreover, it would be useful to constrain the heat capacities together with the adiabatic index, giving a possible explanation of the role played by the temperature.

\section*{Acknowledgements}

P.K.S.D. and L.R. thank the National Research Foundation (NRF) for financial support.


\begin{thebibliography}{99}

\bibitem{galaxy}
S. Capozziello, M. De Laurentis, O. Luongo, A. C. Ruggeri, Galaxies, {\bf 1}, 216, (2013); M. Li, X.-D. Li, S. Wang, Y. Wang, Front. of Phys., {\bf 8}, 828, (2013); P. J. E. Peebles, B. Ratra, Rev. Mod. Phys., {\bf 75}, 559, (2003).

\bibitem{bamba}
K. Bamba, S. Capozziello, S. Nojiri, S. D. Odintsov, Astroph. Sp. Sci., {\bf 342}, 155, (2012).

\bibitem{adam}
A. G. Riess, {\it et al.}, AJ, {\bf 116}, 1009, (1998); S. Perlmutter, {\it et al.}, ApJ, {\bf 517}, 565, (1999).

\bibitem{negativep}
S. Tsujikawa, ArXiv[astro-ph]:1004.1493, \emph{Dark energy: investigation and modeling}, (2010).

\bibitem{lambda1}
C. P. Burgess, ArXiv[hep-th]:1309.4133, \emph{The Cosmological Constant Problem: Why it's hard to get
Dark Energy from Micro-physics}, (2013).

\bibitem{lambda2}
P. J. E. Peebles, B. Ratra, Rev. Mod. Phys., {\bf 75}, 559, (2003).

\bibitem{emay12}
S. D. P. Vitenti and M. Penna-Lima, JCAP, {\bf 09}, 045, (2015); C. Rubano, P. Scudellaro, Gen. Rel. Grav., {\bf 34}, 1931,
(2002); M. Kunz, Phys. Rev. D, {\bf 80}, 123001, (2009); O. Luongo, G. B. Pisani, A. Troisi, \emph{}, ArXiv[gr-qc]: 1512.07076, (2015).

\bibitem{lambda3}
S. Weinberg, Rev. Mod. Phys., {\bf 61}, 1, (1989)

\bibitem{eoevolve}
A. R. Liddle, P. Mukherjee, D. Parkinson, Y. Wang, Phys. Rev. D, {\bf 74}, 123506, (2006); J. Park, C.-G. Park, J.-C. Hwang, Phys. Rev. D, {\bf 84}, 023506, (2011); A. Upadhye, J. Kwan, A. Pope, K. Heitmann, S. Habib, H. Finkel, N. Frontiere, Phys. Rev. D, {\bf 93}, 063515, (2016); R. Nair, S. Jhingan, Jour. Cosm. Astrop. Phys., {\bf 02}, 049, (2013).

\bibitem{revcop}
J. E. Copeland, M. Sami, S. Tsujikawa, Int. J. Mod. Phys. D, {\bf 15}, 1753, (2006).

\bibitem{thermodynamic}
T. Padmanabhan, Current Sci., {\bf 109}, 2236, (2015); G. W. Gibbons, S. W. Hawking, Phys. Rev. D, {\bf 15}, 2738, (1977); H. Quevedo, R. Sussman, Class. Quant. Grav., {\bf 12}, 859, (1995).

\bibitem{revtermo}
H. Quevedo, J. Math. Phys., {\bf 48}, 013506, (2007).


\bibitem{razi1}
S. Viaggiu, Gen. Rel. Gravit., {\bf 47}, 8, (2015); Y. Gong, A. Wang, Phys. Rev. Lett., {\bf 99}, 211301, (2007); M. Akbar, R.-G. Cai, Phys. Rev. D, {\bf 75}, 084003, (2007); R.-G. Cai, S. P. Kim, JHEP {\bf 050}, 0502, (2005); H. H. B. Silva, R. Silva, R. S. Goncalves, Z.-H. Zhu, J. S. Alcaniz, Phys. Rev. D, {\bf 88}, 127302, (2013); R. Silva, R. S. Goncalves, J. S. Alcaniz, H. H. B. Silva, Astr. and Astroph., {\bf 537}, A11, (2012).

\bibitem{altro}
R. Silva, J. S. Alcaniz, Phys. Lett. A, {\bf 313}, 393, (2003); A. Aviles, N. Cruz, J. Klapp, O. Luongo, Gen. Rel. Grav., {\bf 47}, 5, 63, pp. 20, (2015); A. Aviles, J. L. Cervantes-Cota, J. Klapp, O. Luongo, H. Quevedo in \emph{Selected Topics of Computational and Experimental Fluid Mechanics' Springer Book Series: Environmental Science and Engineering}, ArXiv[physics]:1502.05661; B. Einarsson, Phys. Lett. A, {\bf 332}, 335, (2004).

\bibitem{ioequevedo}
Orlando Luongo, Hernando Quevedo, Gen. Rel. Grav., {\bf 46}, 1649,  8, (2014).

\bibitem{alca}
J. A. S. Lima, J. S. Alcaniz, Astron. Astrophys., {\bf 348}, 1-7, (1999).


\bibitem{bll}
H. H. B. Silva, R. Silva, R. S. Goncalves, Z.-H. Zhu, J. S. Alcaniz, Phys. Rev. D,{\bf 88}, 127302, (2013); R. Silva, R. S. Goncalves, J. S. Alcaniz, H. H. B. Silva, Astr. and Astroph., {\bf 537}, A11, (2012); R. G. Cai, S. P. Kim, JHEP, 0502, {\bf 050}, (2005).

\bibitem{herny}
A. Krasinski, H. Quevedo, R. Sussman, J. Math. Phys., {\bf 38}, 2602, (1997).

\bibitem{boul}
D. Lynden-Bell, R. Wood, Mon. Not. Roy. Astr., Soc., {\bf 138}, 495, (1968); D. Lynden-Bell, Physica A, {\bf 263}, 1, (1999); T. Padmanabhan, Phys. Rep., {\bf 188}, 285, (1990); B. Einarsson, Phys. Lett. A, {\bf 332}, 335, (2004).

\bibitem{young}
Y. S. Myung, Phys.Lett.B, {\bf 671}, 216, (2009).

\bibitem{horizon}
R. G. Cai, L. M. Cao, Y. P. Hu, Class. Quant. Grav., {\bf 26}, 155018, (2009); S. del Campo, I. Duran, R. Herrera, D. Pavon, Phys. Rev. D, {\bf 86}, 083509, (2012).

\bibitem{cosmo1}
S. Weinberg, \emph{Gravitation and Cosmology: Principles and Applications of the General Theory of Relativity}, Wiley, Hoboken, NJ, USA, (1972); M. Visser, Phys. Rev. D, {\bf 56}, 7578, (1997); M. Visser, Science, {\bf 276}, 88, (1997); E. R. Harrison, Nature, {\bf 260}, 591, (1976); V. Sahni, T. D. Saini, A. A. Starobinsky, U. Alam, JETP Lett., {\bf 77}, 201-206, (2003); Pisma Zh. Eksp. Teor. Fiz., {\bf 77}, 249-253, (2003); M. Visser, Gen. Rel. Grav., {\bf 37}, 1541-1548, (2005); C. Cattoen, M. Visser, Class. Quant. Grav., {\bf 25}, 165013, (2008); C. Cattoen, Class. Quant. Grav., {\bf 24}, 5985-5998, (2007); A. Aviles, C. Gruber, O. Luongo, H. Quevedo, Phys. Rev. D, {\bf 86}, 123516, (2012); C. Cattoen, M. Visser, Class. Quant. Grav., {\bf 22}, 4913-4930, (2005).


\bibitem{cosmo4}
P. K. S. Dunsby, O. Luongo, Int. J. Geom. Meth. Mod. Phys. {\bf 13}, 03, 1630002, (2016); O. Luongo, Mod. Phys. Lett. A, {\bf 26}, 1459, (2011); S. Weinberg, \emph{Cosmology}, (Oxford Univ. Press, Oxford), (2008).

\bibitem{cosmo5}
C. Gruber, O. Luongo, Phys. Rev. D, {\bf 89}, 103506, (2014); M. Visser, Class. Quant. Grav., {\bf 21}, 2603, (2004); A. Aviles, A. Bravetti, S. Capozziello, O. Luongo, Phys. Rev. D, {\bf 90}, 043531, (2014); T. D. Saini, S. Raychaudhury, V. Sahni, A. A. Starobinsky, Phys. Rev. Lett., {\bf 85}, 1162, (2000).

\bibitem{transition}
O. Farooq and B. Ratra, Astrophys. J. Lett., {\bf 766}, L7, (2013); S. Capozziello, O. Farooq, O. Luongo, B. Ratra, Phys. Rev. D, {\bf 90}, 044016, (2014).






\bibitem{ref:dark_fluid}
A. Arbey, Phys.Rev. D, {\bf 74}, 043516, (2006); L. Xu, Y. Wang, H. Noh, Phys. Rev. D, {\bf 85}, 043003, (2012); M. Nouri-Zonoz, J. Koohbor, H. Ramezani-Aval, Phys. Rev. D, {\bf 91}, 063010, (2015); W. S. Hipolito-Ricaldi, H.E.S. Velten, W. Zimdahl, JCAP, {\bf 016}, 0906, (2009); I. Brevik, E. Elizalde, O. Gorbunova, A. V. Timoshkin, Eur. Phys. J. C, {\bf 52}, 223, (2007).


\bibitem{ref:union2.1}
N. Suzuki, et al., Astrophys. J., {\bf 85}, 746, (2012); R. Amanullah, et al., Astrophys. J., {\bf 716}, 712, (2010).

\bibitem{ref:Luongo_H(z)}
O. Farooq, B. Ratra, Phys. Lett. B, {\bf 723}, 1, (2013); O. Farooq, B. Ratra, ApJ, {\bf 766}, 7, (2013).

\bibitem{Beutler:2011hx}
  F.~Beutler {\it et al.},  Mon. Not. Roy. Astron. Soc., {\bf 416}, 3017, (2011).

\bibitem{Anderson:2013zyy}
 L.~Anderson {\it et al.} [BOSS Collaboration], Mon. Not. Roy. Astron. Soc., {\bf 441}, 1, 24, (2014).

\bibitem{Blake:2011en}
  C.~Blake {\it et al.},  Mon. Not. Roy. Astron. Soc., {\bf 418}, 1707, (2011).

\bibitem{Bennett:2012zja}
  C.~L.~Bennett {\it et al.} [WMAP Collaboration], Astrophys. J. Suppl.,  {\bf 208}, 20, (2013).

\bibitem{Ade:2015xua}
P. A. R. Ade, et al., [Planck Collaboration], Astron. Astrophys. A, {\bf 16}, 571, (2014).



\bibitem{AIC}
H. Akaike, IEEE Trans. Automat. Control, {\bf 19}, 716, (2006); H. Akaike, J. Econometr., {\bf 16}, 3, (2006).


\bibitem{do}
M. Biesiada, JCAP, {\bf 02}, 003, (2007); W. Godlowski, M. Szydlowski, Phys. Lett. B, {\bf 623}, 10, (2005); M. Szydlowski, W. Godlowski, Phys. Lett. B, {\bf 633}, 427, (2006); M. Szydlowski, A. Kurek, A. Krawiec, Phys. Lett. B, {\bf 642}, 171, (2006); M. Szydlowski, A. Kurek, AIP Conf. Proc., {\bf 861}, 1031, (2006).

\bibitem{tro}
K. P. Burnham, D. R. Anderson, \emph{Model Selection and Multimodel Inference}, New York, Springer, (2002); G. Schwarz, Ann. Stat., {\bf 6}, 461, (1978); M. Kunz, R. Trotta and D. Parkinson, Phys. Rev. D, {\bf 74}, 023503, (2006); M. Li, X. Li, S. Wang, X. Zhang, JCAP, {\bf 036}, 0906, (2009).

\bibitem{quqq}
N. Sugiura, Commun. Stat. Theor. Meth. A, {\bf 7}, 13, (1978).


\end{thebibliography}
\end{document}